# Artificial 'spin ice' in a geometrically frustrated lattice of nanoscale ferromagnetic islands

R. F. Wang[1], C. Nisoli[1], R. S. Freitas[1], J. Li [1], W. McConville[1], B. J. Cooley[1], M. S. Lund[2], N. Samarth[1], C. Leighton[2], V. H. Crespi[1] & P. Schiffer[1]

[1]*Department of Physics and Materials Research Institute, Pennsylvania State University, University Park, Pennsylvania 16802, USA.*

[2]*Department of Chemical Engineering and Materials Science, University of Minnesota, Minneapolis, Minnesota 55455, USA.*

**Frustration, defined as a competition between interactions such that not all of them can be satisfied, is important in systems ranging from neural networks to structural glasses. Geometrical frustration, which arises from the topology of a well-ordered structure rather than from disorder, has recently become a topic of considerable interest[1]. In particular, geometrical frustration among spins in magnetic materials can lead to exotic low-temperature states[2], including 'spin ice', in which the local moments mimic the frustration of hydrogen ion positions in frozen water[3–6]. Here we report an artificial geometrically frustrated magnet based on an array of lithographically fabricated single-domain ferromagnetic islands. The islands are arranged such that the dipole interactions create a two-dimensional analogue to spin ice. Images of the magnetic moments of individual elements in this correlated system allow us to study the local accommodation of frustration. We see both ice-like short-range correlations and an absence of long-range correlations, behaviour which is strikingly similar to the low-temperature state of spin ice. These results demonstrate that artificial frustrated magnets can provide an uncharted arena in which the physics of frustration can be directly visualized.**

In one of the most common frustrated systems, ordinary water ice, hydrogen ions follow the so-called 'ice rules'. These rules require that the four hydrogens surrounding each oxygen atom be placed in a tetrahedral coordination such that two are close to the central oxygen atom, while the other two are closer to neighbouring oxygen atoms[7]. The spin ice materials have the pyrochlore structure in which magnetic rare-earth ions form a lattice of corner-sharing tetrahedra. In these materials, the ice rules are manifested by a minimization of the spin–spin interaction energy when two spins point inward and two spins point outward on each tetrahedron. At low temperatures, the spins freeze into an exotic disordered state that has many of the hallmarks of glassiness[3–5]. This ice-like state is different from the disorder-based spin glasses, in that it is associated with a very narrow range of spin relaxation times due to the well-ordered lattice[8,9], and it is also quite different from spin liquid states seen in other geometrically frustrated rare-earth magnets[10,11], in that the spins do give evidence of freezing into a static configuration at the lowest temperatures.

One of the most fascinating aspects of geometrically frustrated magnets is how the spins locally accommodate the frustration of the spin–spin interactions. As a practical matter, however, individual spins within a material are difficult to probe experimentally without altering the state of the system. One way around this problem is to create a frustrated system in which the individual elements can be directly probed. Toward this end, previous workers have fabricated arrays of superconducting rings or Josephson junctions, in which the interacting moments are trapped flux quanta created by the application of a magnetic field at low temperatures[12,13]. A much closer analogy to the frustrated magnetic materials can be provided, however, by arrays of interacting single-domain ferromagnetic islands in which the moments are intrinsic (that is, they do not require the application of an external field). Advances in lithography allow great flexibility in the design of ferromagnetic island arrays, and such arrays have the added advantage of being accessible at room temperature. Studies of lines and rectangular arrays of such ferromagnetic islands show that pairwise dipolar interactions between them can be significant[14–19]. These results suggest that frustration effects should be important if a lattice of such islands can be fabricated with a frustrated geometry.

We studied frustrated arrays consisting of two-dimensional square lattices of elongated permalloy islands (shown schematically in Fig. 1a) with the long axes of the islands alternating in orientation along the two principal directions of the array lattice (fabrication details are given in the Methods section). We studied arrays with lattice parameters ranging from 320 nm to 880 nm, with a fixed island size (80 nm×220 nm laterally and 25 nm thick). This size is sufficiently small that the atomic spins were ferromagnetically aligned in a single domain (as demonstrated below), but large enough that the moment configuration was stable at 300 K. The moment of each island was approximately $3×10^7$ Bohr magnetons (estimated from the known magnetization of permalloy), and the magnetic field from an island was of the order of 10 Oe at the centre of nearest neighbour islands—leading to an interaction energy of the order of $10^{-19}$ Joule, equivalent to $10^4$ K, between nearest neighbours (the exact value naturally depends on the lattice spacing). The magnetocrystalline anisotropy of permalloy is effectively zero, so that the shape anisotropy of each island (the self-energy of the island's magnetic moment, which is controlled by its shape) forced its magnetic moment to align along the long axis, thus making the islands effectively Ising-like. Finite element modelling of the islands using the OOMMF code[20] indicated that the magnetic fields originating from other islands did not substantially alter this Ising-like behaviour.

The intrinsic frustration on this lattice is similar to that in the 'square ice' model (see ref. 21 and references therein), and can best be seen by considering a vertex at which four islands meet. Nearest-neighbour islands on such a vertex are perpendicular to each other, while the second nearest neighbour islands are collinear and directly across the vertex from each other (see Fig. 1b). A pair of moments on a vertex can be aligned either to maximize or to minimize the dipole interaction energy of the pair. As shown in the figure, it is energetically favourable when the moments of the pair of islands are aligned so that one is pointing into the centre of the vertex and the other is pointing out, while it is energetically unfavourable when both moments are pointing inward or both are pointing outward. For the vertex as a whole, there are four distinct topologies for the configurations of the four moments with a total multiplicity of 16, as shown in Fig. 1c. We label the configurations I–IV in the order of increasing magnetostatic energy, but no configuration can minimize all of the dipole–dipole interactions (even type I only minimizes the energy for four of the six pairs in a vertex), and thus the system is frustrated.

The lowest energy vertex configurations (I and II) have two of the moments pointing in toward the centre of the vertex, and two pointing out. Although the interactions between all pairs of spins on the vertex are not equivalent, these energetics are analogous to the two-in/two-out ice rules for the atomic moments on a tetrahedron in spin ice. For arrays with a lattice constant of 320 nm, the energy difference between vertices of types I and III is more than twice as large as the energy difference between vertices of types I and II, and the energy difference between types I and IV is more than six times as large (based on OOMMF calculations of relaxed magnetostatic energies). The two-in/two-out motifs (types I and II) therefore dominate within a large manifold of closely spaced low-energy magnetic states. Topological considerations further favour the creation of magnetic states that are dominated by frustrated mixtures of types I and II. For example, a domain boundary between regions of types I and II is essentially seamless, requiring no vertices of types III or IV. The situation contrasts sharply with that of a traditional Ising ferromagnet or antiferromagnet, wherein magnetic domain walls contain highly unfavourable anti-aligned spin pairs.

Magnetic force microscopy (MFM) allowed us to image the orientations of all of the moments in a large area (10 μm×10 μm), far from the edges of the arrays. To enable the system to settle into a low energy configuration, we followed a protocol developed by previous authors[16,18] and rotated the samples in a magnetic field which decreased stepwise from above to below the coercive field. MFM images of the system after such field treatment revealed no measurable residual magnetic moment for the array, and a ten-fold variation of the step dwell times did not significantly alter the distribution of vertex types described below.

In Fig. 2 we show an atomic force microscope (AFM) image and an MFM image of a portion of a typical array. The black and white spots in the MFM image, which indicate the north and south poles of the ferromagnetic islands, confirm the single-domain nature of the islands and demonstrate the dominance of the shape anisotropy in aligning the magnetization of each island along its long axis. From the MFM data, we can easily determine the moment configuration of the array (as indicated by the arrows in Fig. 1a). These data demonstrate that the many vertex types anticipated in Fig. 1c can be directly observed in the actual system. In order to probe the nature of frustration in this system, we studied how the properties varied with the spacing between the islands, counting between 1,000 and 3,000 islands in measurements of 2–4 different arrays for each lattice spacing. This allowed us direct control over the frustrated interactions, something which is not easily attainable in geometrically frustrated magnetic materials.

An immediate question is whether our arrays obeyed the ice rules—that is, did a preponderance of the vertices fall into a two-in/two-out configuration (type I or II)? By simple counting arguments (see Fig. 1c) we can predict the expected distribution of different vertex types if the moments were non-interacting and randomly oriented. One would expect only 37.5% of the vertices to have a two-in/two-out configuration if the orientations were random; an excess of such vertices would indicate that interactions are determining the moment configuration. We compute the excess percentage for each type of vertex, defined as the difference between the percentage observed and that expected for a random distribution. We plot this excess versus lattice spacing in Fig. 3a for each of the four vertex types, as well as for types I and II combined. The excess percentage of vertices with a two-in/two-out configuration (types I and II) was well over 30% for the smallest lattice spacing; in other words, over 70% of all vertices had a spin-ice-like configuration. This excess percentage decreased monotonically with increasing lattice spacing (decreasing interactions), approaching zero for our largest lattice spacing, as would be expected for non-interacting (randomly oriented) moments. In fact, the excess for all vertex types approached zero as the lattice spacing increased, lending credence both to our understanding of the system and to the effectiveness of the rotating-field method in enabling facile local re-orientation of the moments.

To further understand the nature of frustration in this system, we also studied the pairwise correlations between the Ising-like moments of the islands. Defining a correlation function is somewhat complicated by the anisotropic nature of our lattice and that of the dipole interaction. We thus define a set of correlation functions between distinct types of neighbouring pairs. The closest pairing is labelled 'NN' for the nearest neighbour; 'L' denotes the next nearest neighbour pairing, which is in the longitudinal direction of the island; and 'T' denotes the next nearest neighbour in the transverse direction from the island (see Fig. 3b inset). We define a correlation, $C$, such that $C=+1$ if two moments are aligned to minimize the dipole interaction energy, and $C=-1$ if two moments are aligned to maximize the dipole energy. In this way, if the moments for a particular type of neighbouring pair were uncorrelated on the lattice, the average value of $C$ would be zero.

We find that the island pairs which were further separated than the L and T neighbours had weak or zero correlations ($|C|<0.1$) for all lattices. We do see correlations for the NN and T neighbours as shown in Fig. 3b, but somewhat surprisingly, the correlations for

the L neighbours were relatively small. We can understand this as a direct consequence of the frustration in the system. Interaction between the NN neighbours is the strongest, and therefore it is predominant. A pair of islands of type L has a direct interaction (which is somewhat weaker than that of the NN pair), but also an indirect interaction, since the two islands in an L pairing share two NN neighbours. If all of the NN pair energies are minimized, then the L pair energy is maximized, and we believe that this frustration leads to the surprisingly weak correlation between the L neighbours. By the same logic, the relatively strong correlations between the T neighbour pairs also arise from indirect interactions via NN intermediaries. In the case of the T neighbour pairs, if the NN neighbour pair interaction energy is minimized, the indirect interaction energy will also be minimized, and thus the combined effect is to increase correlations as we observe. For all of the neighbour types, we find that the correlations approached zero for the largest lattice parameters, as expected since the interactions should strongly decrease as the islands are separated.

The existence of only short-range order and ice-like correlations on the lattice is precisely analogous to the behaviour of the spin ice materials, in which there is also no experimental evidence for long-range order, only ice-like short-range correlations. While there are theoretical long-range ordered low-energy states for spins on either our lattice or the pyrochlore spin ice lattice[22], the complex energy landscape associated with the frustration leads to a disordered state when thermal or magnetic-field-induced excitations are removed. This is in sharp contrast with unfrustrated lines of ferromagnetic islands, in which longer-range correlations are observed[16,17]. It is interesting that the relative populations of different types of vertices reaches the randomly oriented limit rapidly as the lattice constant increases, and that even within the regime of closely-spaced and therefore strongly interacting islands, the system can access a very wide range of nearly degenerate states. This wide range of accessible states has the potential for importance to applications, since, if information were encoded within a low energy configuration of the moments, the energetic driving force for local magnetization reversals could be suppressed by this near-degeneracy, even for highly dense arrays.

Our demonstration of an artificial frustrated magnet opens the door to a new mode of research wherein a frustrated system can be designed rather than discovered. Future studies could examine a wide range of accessible lattice geometries, rationally designed defect structures, and the effects of dynamic and static applied magnetic field[23]. In addition, the capability to locally probe the magnetic moments, the accessibility at room temperature, and the similarity to patterned magnetic recording media all combine to suggest the potential for novel technological applications that exploit the fundamental nature of frustration.

## Methods

We fabricated the arrays on Si substrates with a native oxide layer, using films of permalloy ($Ni_{0.81}Fe_{0.19}$) with grain size of about 5 nm. We employed a lift-off technique using a polymethylglutarimide (PMGI) and polymethyl methacrylate (PMMA) double layer resist[24,25]. After electron beam exposure, we used methyl isobutyl ketone: isopropyl alcohol (in the ratio of 1:3) to remove exposed PMMA resist, followed by removal of that PMGI resist which is not covered by PMMA. Then we used a molecular beam epitaxy system to grow a 25-nm-thick permalloy film on the pattern at a deposition rate of 0.1 Å s$^{-1}$ at ambient temperature. The permalloy was capped with 3 nm of Al to prevent oxidation of the magnetic material. After a lift-off process in acetone, the PMGI and PMMA resists were removed and the nanometre-scale islands stood on the Si substrate. The total array size ranges from 64 μm×64 μm to 176 μm×176 μm, with the size increasing for less dense arrays (there were 80,000 islands in each array).

We used a Veeco Multimode MFM to detect individual island magnetization under zero magnetic field. Repeated scans demonstrated that the tip did not change the orientation of the island moments. Before measurement, the sample was rotated at 1,000 r.p.m. in an in-plane magnetic field, with the magnetic field starting at 1,300 Oe (well above the coercive field of the islands) and gradually stepping down in magnitude to zero.

**Acknowledgements** We acknowledge financial support from the Army Research Office and the National Science Foundation MRSEC programme, and discussions with P. Crowell and P. Lammert. R.S.F. thanks the CNPq-Brazil for sponsorship.

**Author Information** Reprints and permissions information is available at npg.nature.com/reprintsandpermissions. The authors declare no competing financial interests. Correspondence and requests for materials should be addressed to P.S. (schiffer@phys.psu.edu).

## References

1. Ramirez, A. P. in *Handbook of Magnetic Materials* Vol. 13 (ed. Buschow, K. J. H) 423–520 (Elsevier Science, Amsterdam, 2001).

2. Moessner, R. Magnets with strong geometric frustration. *Can. J. Phys.* **79,** 1283–1294 (2001).

3. Harris, M. J., Bramwell, S. T., McMorrow, D. F., Zeiske, T. & Godfrey, K.W. Geometrical frustration in the ferromagnetic pyrochlore $Ho_2Ti_2O_7$. *Phys. Rev. Lett.* **79,** 2554–2557 (1997).

4. Siddharthan, R. *et al*. Ising pyrochlore magnets: low-temperature properties, "ice rules," and beyond. *Phys. Rev. Lett.* **83,** 1854–1857 (1999).

5. Ramirez, A. P., Hayashi, A., Cava, R. J., Siddharthan, R. & Shastry, B. S. Zero-point entropy in 'spin ice'. *Nature* **399,** 333–335 (1999).

6. Bramwell, S. T. & Gingras, M. J. P. Spin ice state in frustrated magnetic pyrochlore materials. *Science* **294,** 1495–1501 (2001).

7. Pauling, L. *The Nature of the Chemical Bond* 301–304 (Cornell Univ. Press, Ithaca, New York, 1945).

8. Snyder, J., Slusky, J. S., Cava, R. J. & Schiffer, P. How 'spin ice' freezes. *Nature* **413,** 48–51 (2001).

9. Snyder, J. *et al*. Low-temperature spin freezing in the $Dy_2Ti_2O_7$ spin ice. *Phys. Rev. B* **69,** 064414 (2004).

10. Tsui, Y. K., Burns, C. A., Snyder, J. & Schiffer, P. Magnetic field induced transitions from spin glass to liquid to long range order in a 3D geometrically frustrated magnet. *Phys. Rev. Lett.* **82,** 3532–3535 (1999).

11. Gardner, J. S. *et al*. Cooperative paramagnetism in the geometrically frustrated pyrochlore antiferromagnet $Tb_2Ti_2O_7$. *Phys. Rev. Lett.* **82,** 1012–1015 (1999).

12. Davidovic, D. *et al*. Correlations and disorder in arrays of magnetically coupled superconducting rings. *Phys. Rev. Lett.* **76,** 815–818 (1996).

13. Hilgenkamp, H. *et al*. Ordering and manipulation of the magnetic moments in large-scale superconducting π-loop arrays. *Nature* **422,** 50–53 (2003).

14. Cowburn, R. P. & Welland, M. E. Room temperature magnetic quantum cellular automata. *Science* **287,** 1466–1468 (2000).

15. Ross, C. A. *et al*. Magnetic behavior of lithographically patterned particle arrays (invited). *J. Appl. Phys.* **91,** 6848–6853 (2002).

16. Cowburn, R. P. Probing antiferromagnetic coupling between nanomagnets. *Phys. Rev.* B **65,** 092409 (2002).

17. Martin, J. I., Nogues, J., Liu, K., Vicent, J. L. & Schuller, I. K. Ordered magnetic nanostructures: fabrication and properties. *J. Magn. Magn. Mater.* **256,** 449–501 (2003).

18. Imrea, A., Csabaa, G., Bernstein, G. H., Porod, W. & Metlushko, V. Investigation of shape-dependent switching of coupled nanomagnets. *Superlatt. Microstruct.* **34,** 513–518 (2003).

19. Stamps, R. L. & Camley, R. E. Frustration and finite size effects of magnetic dot arrays. *J. Magn. Magn. Mater.* **177,** 813–814 (1998).

20. The Object Oriented MicroMagnetic Framework (OOMMF) project at ITL/NIST. <http://math.nist.gov/oommf/>.

21. Lieb, E. H. & Wu, F. Y. in *Phase Transitions and Critical Phenomena* (eds Domb, C. & Green, M. S.) Vol. I (Academic, London, 1972).

22. Melko, R. G., den Hertog, B. C. & Gingras, M. J. P. Long-range order at low temperatures in dipolar spin ice. *Phys. Rev. Lett.* **87,** 067203 (2001).

23. Zhitomirsky, M. E., Honecker, A. & Petrenko, O. A. Field induced ordering in highly frustrated antiferromagnets. *Phys. Rev. Lett.* **85,** 3269–3272 (2001).

24. Yang, X. M. *et al*. Fabrication of sub-50nm critical feature for magnetic recording device using electron-beam lithography. *J. Vac. Sci. Technol. B* **21,** 3017–3020 (2003).

25. Lin, C. K., Wang, W., Hwu, H. & Chan, Y. Single step electron-beam lithography for asymmetric recess and gamma gate in high electron mobility transistor fabrication. *J. Vac. Sci. Technol. B*. **22,** 1723–1726 (2004).

**Figure Captions**

**Figure 1 Illustration of frustration on the square lattice used in these experiments.** Each island in the lattice is a single-domain ferromagnet with its moment pointing along the long axis, as indicated by the arrow. **a**, The geometry of the lattice studied. The arrows indicate the directions of moments corresponding to the MFM image of Fig. 2b. **b**, Vertices of the lattice with pairs of moments indicated, illustrating energetically favourable and unfavourable dipole interactions between the pairs. **c**, The 16 possible moment configurations on a vertex of four islands, separated into four topological types. The percentages indicate the expected fraction of each type if the individual moment orientations on an array were completely random.

**Figure 2 AFM and MFM images of a frustrated lattice. a**, An AFM image of a typical permalloy array with lattice spacing of 400 nm. **b**, An MFM image taken from the same array. Note the single-domain character of the islands, as indicated by the division of each island into black and white halves. The moment configuration of the MFM image is illustrated in Fig. 1a. The coloured outlines indicate examples of vertices of types I, II and III in pink, blue and green respectively.

**Figure 3 Statistics of moment configurations.** These statistics were obtained from between 1,000 and 3,000 islands in combined measurements of 2–4 different arrays for each lattice spacing. **a**, The excess percentages of different vertex types, plotted as a function of the lattice spacing of the underlying square array lattice. Note that the excess percentages approach zero for the largest lattice spacing. **b**, The correlations between different pairs of the islands as a function of the lattice spacing of the underlying square lattice. The inset shows our definitions of the near neighbour pairs from the grey central island (see text for details). For both the correlations and the vertex statistics, the typical variation between images for nominally identically prepared samples was <10% for the closely spaced lattices in which we had more than 1,000 islands in a single image, but up to 50% for the more widely spaced lattices in which we had only a few hundred islands per image.

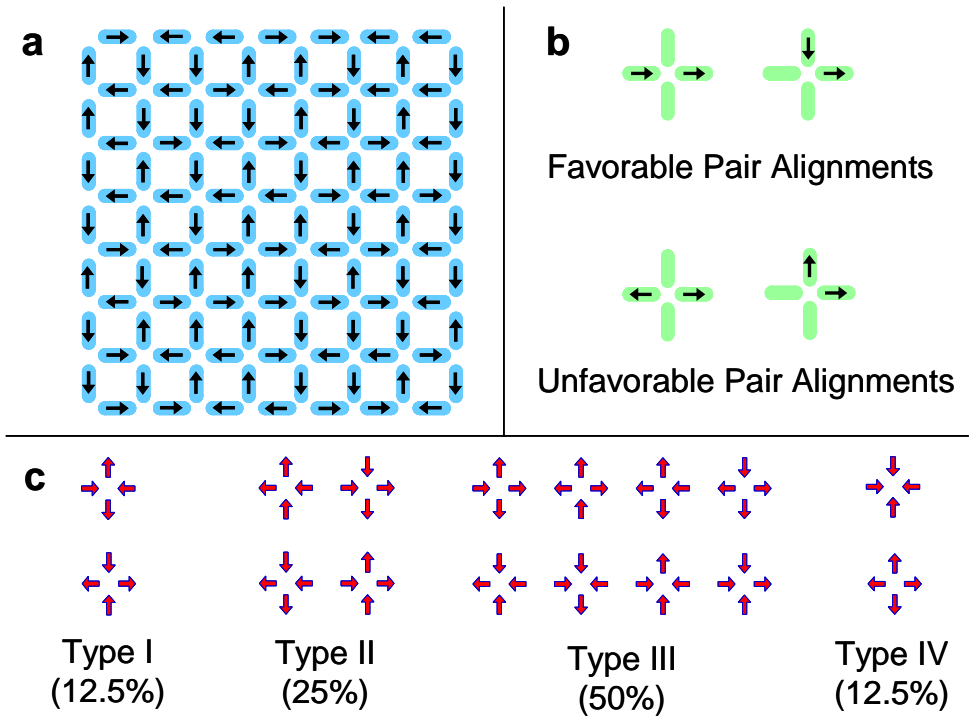

**a**

**b**

Favorable Pair Alignments

Unfavorable Pair Alignments

**c**

Type I
(12.5%)

Type II
(25%)

Type III
(50%)

Type IV
(12.5%)

Figure 1.  Wang *et al.*

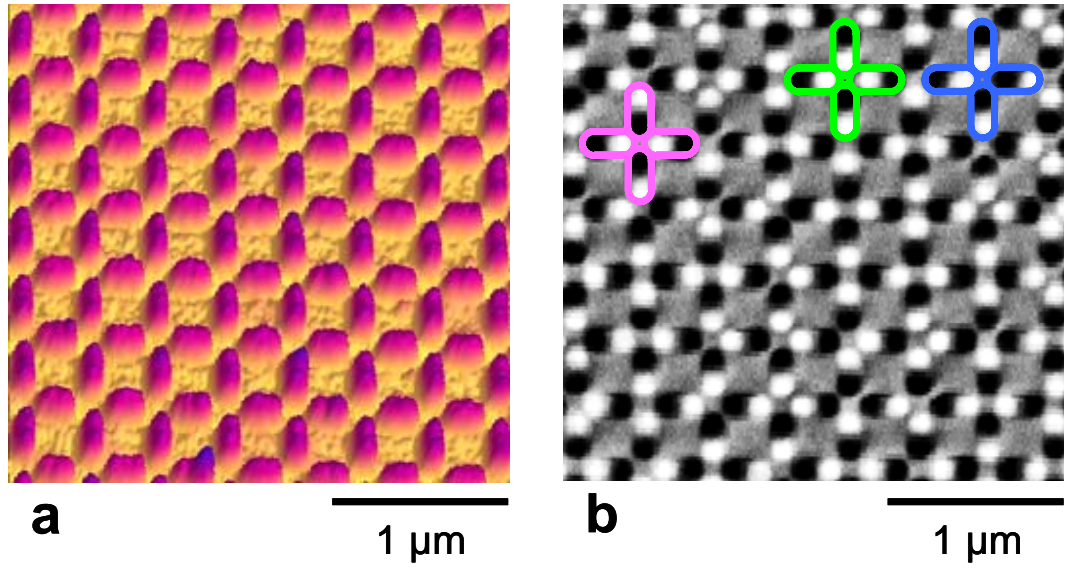

Figure 2.  Wang *et al.*

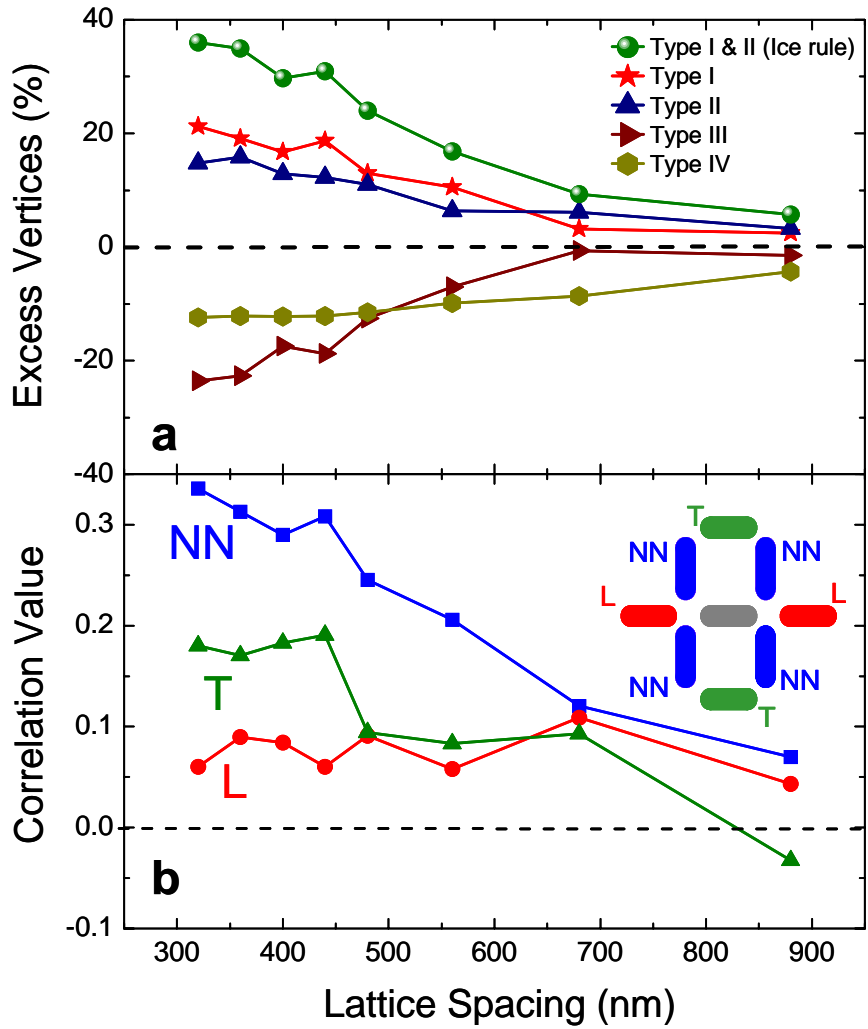

Figure 3.  Wang *et al.*